\documentclass{ws-procs9x6} 

\begin{document}

\pagestyle{empty}

\title{Glueball as a bound state in the self-dual homogeneous vacuum gluon field
\footnote{\uppercase{T}his work is supported by 
\uppercase{R}\uppercase{F}\uppercase{F}\uppercase{I} grants 02--02--27064, 01--02--17200}}

\author{\underline{Ja.~V. BURDANOV}}

\address{Laboratory of Nuclear Problems, JINR, Russia, \\
E-mail: burdanov@thsun1.jinr.ru}

\author{G.~V. EFIMOV}

\address{
Bogoliubov Laboratory of Theoretical Physics, JINR, Russia \\
E-mail: efimovg@thsun1.jinr.ru}  

\maketitle

\abstracts{
Using a simple relativistic QFT model of scalar fields
we demonstrate that the analytic confinement (propagator is an
entire function in the complex $p^2$--plane) and the weak
coupling constant lead to the Regge behaviour of the
two-particle bound states.
In QCD we assume that the gluon vacuum is realized by the self-dual
homogeneous classical field which is the solution of the Yang-Mills
equations. This assumption leads to analytical confinement
of quarks and gluons. We extract the colorless $0^{++}$ two-gluon
state from the QCD generating functional in the one-gluon exchange
approximation.
The mass of this bound state is defined by the Bethe-Salpeter equation.
The glueball mass is $1765~{\rm MeV}$ for $\alpha_s=0.33$
if the gluon condensate is
$\langle (\alpha_s/\pi) G G \rangle=0.012~{\rm GeV}^4$.}

\section{Virton model}

In this section we demonstrate that the analytic confinement and Bethe-Salpeter (BS) 
equation in weak coupling regime lead to the pure Regge spectrum  
for all bound states with radial and orbital quantum numbers.

As an example we use the virton model of scalar field\cite{efim1}.
The equation for the free field $\varphi(x)$ looks like:
$\Lambda^2~e^{{p^2\over \Lambda^2}}\tilde{\varphi}(p)=0$ and 
the solution $\varphi(x)\equiv0$ because the function
$e^{{p^2\over \Lambda^2}}\neq0$, i.e., it has no zeroes for any
real or complex $p^2$. 
Exactly this property means {\it analytic confinement}. 
The field $\varphi(x)$ can exist in the virtual
state only, so that it can be called virton field\cite{efim2}. 

The propagator $\Lambda^2 \tilde D(p^2)=e^{-{p^2\over\Lambda^2}}$
is an entire analytical function in the complex $p^2$--plane,
i.e., there are no poles corresponding to free particles.
It guarantees the confinement of ``particles'' $\varphi(x)$
in each perturbation order over the dimensionless coupling
constant $\lambda=\left({3g\over 4\pi \Lambda}\right)^2\leq 1$.
The parameter $\Lambda$ characterizes the scale of the confinement region.

The mass $M_Q$ of the bound states with quantum numbers $Q=(n,l)$ 
is defined by the BS equation in one-boson exchange approximation\cite{efim3}.
The virton model gives the pure Regge behaviour 
\begin{eqnarray*}
M^2_{nl}&=&2\Lambda^2\ln\left(\lambda_c/\lambda\right)+
(l+2n)\Lambda^2\ln(2\lambda_c).
\end{eqnarray*}
We can see that bound states exist for small coupling constant
$\lambda < \lambda_c\sim 7$ and their masses grow when $\lambda\to0$ as
$M_Q\sim \Lambda\sqrt{2\ln\left(\lambda_c/\lambda\right)}$.
The slope of the Regge trajectories is defined by the scale of the confinement  
region $\Lambda$ and does not depend on the coupling constant $\lambda$.

\section{Glueball mass}

We suppose\cite{efim4} that the QCD vacuum is realized by a homogeneous 
(anti-) self-dual gluon field 
\begin{eqnarray}
\label{B}
&&B^a_\mu(x)=n^a B b_{\mu\nu}x_\nu,~~b_{\mu\rho}b_{\rho\nu}=-\delta_{\mu\nu},~~
\tilde{b}_{\mu\nu}=\epsilon_{\mu\nu\alpha\beta}b_{\alpha\beta}=\pm b_{\mu\nu},
\end{eqnarray}
which is the solution of the Yang-Mills equations.
The parameter $B$ is a tension of the field.

This assumption leads to full quark and partial gluon ana\-ly\-tic 
con\-fi\-ne\-ment\cite{efim1,efim4}.
It means that in the presence of field $(\ref{B})$ quark and glu\-on pro\-pa\-ga\-tors 
are entire analytical functions in $p^2$ complex plane.

The mechanism of the creation of two-gluon bound states can be described in the following way.
\begin{itemize}
\item Gluon-gluon effective Lagrangian in one-gluon exchange approximation is extracted 
from 3-gluon interaction in Euclidean QCD generating functional  
in the presence of background field.
\item Bilocal colorless scalar gluon currents are derived using the Fierz transformation.
\item Untroducing the orthonormal system of functions, bilocal gluon currents can be decomposed 
into a series of nonlocal scalar colorless currents with radial and orbital quantum numbers.
\item Bosonization procedure of the nonlocal 4-gluon interaction
consists in a choice of orthonormal functions which correspond to amplitudes
of Bethe-Salpeter equation. 
The kernel of BS equation is written in a symmetric hermitian form that allows
one to find the ground state BS amplitude using the variational method. 
The crucial point is that only confined gluons are responsible 
for arising two-gluon bound states.
\item Glueball mass spectrum is determined by BS equation. 
\end{itemize}
The parameter $\Lambda$ can be connected with the gluon vacuum condensate
$
\left\langle {\alpha_s\over\pi}G_{\mu\nu}^aG_{\mu\nu}^a\right\rangle_0=
{4\over\pi^2}(gB)^2={4\over 3\pi^2}\Lambda^4,
$
so that, using the customary value of the gluon condensate\cite{shif} 
we obtain $\Lambda=546~{\rm MeV}$.
Our numerical value of scalar glueball mass for\cite{hin} $\alpha_s=0.33$ is 
$M_{0^{++}}=1765~{\rm MeV}$. 
This result is in reasonable agreement with the lattice data\cite{lattice}.

\section{Conclusion}

In the conclusion we would like to list the main points of
our approach.
\begin{romanlist}
\item Analytic confinement and BS equation in a weak coupling regime 
lead to the Regge spectrum of bound states (see also\cite{efim5}).
\item Self-dual homogeneous vacuum gluon field (\ref{B}) leads
to full quark confinement and partial gluon confinement.
\item Only confined gluon degrees of freedom are responsible 
for glueball bound states.
\item Numerical result for the mass of the $0^{++}$ glueball
is quite reasonable and with the results of\cite{efim4} gives hope
that our main assumption (\ref{B}) contains ``a grain of truth''
about the structure of the QCD vacuum.
\end{romanlist}

\section*{Acknowledgments}
We wish to thank A. Dorokhov, F. Lenz, S. Mikhailov, S. Nedelko  and A. Illarionov
for helpful discussions.


\begin{thebibliography}{99}
\bibitem{efim1} G.V. Efimov and Ja.V. Burdanov,
{\it Phys. Rev.} {\bf D64}, 014001 (2001). 
\bibitem{efim2} G.V. Efimov and M.A. Ivanov, {\it IOP Publishing LTD, London}, (1993).
\bibitem{efim3} G.V. Efimov, hep-ph/9907483 (1999).
\bibitem{efim4} G.V. Efimov {\it et al}, 
{\it Phys. Rev.} {\bf D51}, 174 (1995),
{\it Eur. Phys. J.} {\bf C1} 343 (1998),
{\it Phys. Rev.} {\bf D54} 4483 (1996),
{\it Phys. Rev.} {\bf D59} 014026 (1999).
\bibitem{shif} M.A. Shifman, {\it World Scientific} {\bf 62}, (1999).
\bibitem{hin} I. Hinchliffe, {\it Annu. Rev. Nucl. Part. Sci.} 
{\bf 50}, 643 (2000).
\bibitem{lattice} C. Morningstar and M. Peardon, 
{\it Phys. Rev.} {\bf D60}, 034509 (1999). 
\bibitem{efim5} G.V. Efimov and G. Ganbold,
{\it Phys. Rev.} {\bf D65}, 054012 (2002). 
\end{thebibliography}
\end{document}